\documentclass[pdflatex,sn-mathphys-num]{sn-jnl}

\usepackage{multirow}%
\usepackage{mathrsfs}%
\usepackage[title]{appendix}%

\usepackage{amsmath,amssymb,amsthm}     
\usepackage{algorithm,algorithmic}      
\usepackage{graphicx}                   
\usepackage{subfigure}  
\usepackage{tabularx,booktabs,multirow} 
\usepackage{xcolor}                     
\usepackage{bm}                         
\usepackage{float}
\usepackage{placeins}
\usepackage{bbding}
\usepackage{svg}

\newcommand{\bR}{\mathbb{R}}
\newcommand{\bE}{\mathbb{E}}

\newcommand{\m}[1]{\boldsymbol{#1}}

\newcommand{\cN}{\mathcal{N}}

\renewcommand{\qed}{\hfill \rule{6pt}{6pt}} 

\newcommand{\sbra}[1]{\left(#1\right)} 
\newcommand{\mbra}[1]{\left[#1\right]} 

\begin{document}

\title[Article Title]{DREAM-$\text{B}^3$P: Dual-Stream Transformer Network Enhanced by Feedback Diffusion Model for Blood-Brain Barrier Penetrating Peptide Prediction}

\author[1]{\fnm{Kaijie} \sur{Wang}}\email{kaijiewang@stu.xjtu.edu.cn}

\author[1]{\fnm{Le} \sur{Yin}}\email{yinle0209@stu.xjtu.edu.cn}

\author[1]{\fnm{Aodi} \sur{Tian}}\email{tad\_20021009@163.com}

\author[1]{\fnm{Zhiqiang} \sur{Wei}}\email{zhiqiang.wei@xjtu.edu.cn}

\author*[1]{\fnm{Zai} \sur{Yang}}\email{yangzai@xjtu.edu.cn}

\author[2]{\fnm{Min} \sur{Han}}\email{hanmin@zju.edu.cn}

\author[2]{\fnm{Qichun} \sur{Wei}}\email{qichun\_wei@zju.edu.cn}

\author[1]{\fnm{Sheng} \sur{Wang}}\email{shengwang@xjtu.edu.cn}

\affil[1]{\orgdiv{School of Mathematics and Statistics}, \orgname{Xi'an Jiaotong University}, \orgaddress{\street{No. 28 Xianning West Road}, \city{Xi'an}, \postcode{710049}, \state{Shaanxi}, \country{China}}}

\affil[2]{\orgdiv{School of Medicine}, \orgname{Zhejiang University}, \orgaddress{\street{866 Yuhangtang Road}, \city{Hangzhou}, \postcode{310058}, \state{Zhejiang}, \country{China}}}

\abstract{
\textbf{Introduction:} The blood–brain barrier (BBB) protects the central nervous system but prevents most neurotherapeutics from reaching effective concentrations in the brain. BBB-penetrating peptides (BBBPs) offer a promising strategy for brain drug delivery; however, the scarcity of positive samples and severe class imbalance hinder the reliable identification of BBBPs.

\textbf{Objectives:} Our goal is to alleviate class imbalance in BBBP prediction and to develop an accurate, interpretable classifier for BBBP prediction.
	
\textbf{Methods:} We propose DREAM-B$^3$P, which couples a feedback diffusion model (FB-Diffusion) for data augmentation with a dual-stream Transformer for classification. FB-Diffusion learns the BBBP distribution via iterative denoising and uses an external analyzer to provide feedback, generating high-quality pseudo-BBBPs. The classifier contains a sequence stream that extracts structural features from peptide sequences and a physicochemical stream that captures physicochemical features such as hydrophobic surface area, molecular charge, number of rotatable bonds, and polarizability. Combining the two features leads to superior BBBP predictive performance.
	
\textbf{Results:} On a benchmark test set containing equal numbers of BBBPs and non-BBBPs, DREAM-B$^3$P surpasses baseline methods (Deep-B$^3$P, B$^3$Pred, BBPpredict and Augur), improving AUC/ACC/MCC by 4.3\%/17.8\%/14.9\%, respectively, over the second-best method. 

\textbf{Conclusion:} By integrating feedback diffusion with a dual-stream Transformer classifier, DREAM-B$^3$P effectively mitigates data scarcity and imbalance and achieves state-of-the-art performance.
}


\maketitle
\section{Introduction}\label{sec:introduction}
The blood–brain barrier (BBB) is critical for maintaining cerebral homeostasis and a major obstacle for central nervous system (CNS) drug delivery \cite{pardridge2005blood, sanchez2017blood, tang2022merged}. Composed of brain microvascular endothelial cells, pericytes, and astrocytes within the neurovascular unit, the BBB tightly restricts the entry of exogenous molecules into the brain via tight junctions and selective transport systems \cite{chen2012modern,malakoutikhah2011shuttle,pardridge2012drug}. While this protective barrier effectively safeguards neural tissue, it also prevents numerous candidate therapeutics from reaching effective exposure levels in the brain, posing a common challenge in the development of treatments for neurodegenerative diseases, brain tumors, epilepsy, and other CNS disorders \cite{menken2000global, pangalos2007drug,zhou2021brain}.

Peptides capable of crossing the BBB, known as blood–brain barrier penetrating peptides (BBBPs), offer a promising pathway for brain drug delivery \cite{oller2016blood, kristensen2017routes,diaz2018branched}. These peptides can exploit endogenous transport mechanisms such as receptor-mediated transcytosis and carrier-mediated transport to traverse into the brain parenchyma. They may serve either as therapeutic agents themselves or as ``molecular shuttles" for delivering proteins, oligonucleotides, or nanocarriers. BBBPs can be derived from natural protein fragments or obtained through combinatorial chemistry and sequence design, offering advantages over large-molecule drugs such as ease of synthesis, modifiability, and lower toxicity \cite{batrakova2011cell,bertrand2010transport,bickel2001delivery}. However, high experimental screening costs and long validation cycles have resulted in a severe scarcity of publicly available BBBPs.

To reduce screening and validation costs, researchers have introduced many artificial intelligence methods to facilitate BBBP identification. Early approaches predominantly relied on computational features \cite{xu2021comprehensive,chou2005using,tan2019identification,zou2023accurately,dubchak1995prediction,jin2019dunet,saravanan2015harnessing,yang2023gender,chou2005progress,chou2000prediction}—such as amino acid composition, composition-transition-distribution, dipeptide composition, and physicochemical properties like hydrophobic surface area, molecular charge, number of rotatable bonds, and polarizability—combined with traditional machine learning models including logistic regression, support vector machines, random forests, and gradient boosting \cite{bishop2006pattern,jordan2015machine}, typically achieving competitive performance \cite{dai2021bbppred,kumar2021b3pred,chen2022bbppredict,ma2023prediction}. Due to the scarcity of positive samples, however, these methods are limited by class imbalance, which constrains their generalization.
Subsequently, data augmentation and representation learning techniques were explored. Approaches such as SMOTE and Borderline-SMOTE oversampling, as well as simple peptide sequence perturbations (e.g., random masking or sequence reversal), have been employed. Generative models like the feedback generative adversarial network (FB-GAN) \cite{gupta2019feedback} were also adopted to expand the positive samples, while convolutional networks \cite{krizhevsky2012imagenet,he2016deep}, recurrent neural networks \cite{graves2012long,cho2014learning}, and Transformer \cite{vaswani2017attention,devlin2019bert} were utilized to directly learn sequence features \cite{tang2024deepb3p}. 
Despite these advances, two key challenges remain. First, pseudo-BBBPs generated by existing generative models often suffer from unstable quality and limited diversity, reducing their utility for data augmentation. Second, many deep models tend to overlook either the peptide’s sequence structure or its physicochemical properties, which imposes a performance bottleneck.

To address these issues, we introduce improvements in both data augmentation and classification modeling. Leveraging the stability, diversity and conditional guidance capability of diffusion models \cite{ho2020denoising}—which learn data distributions and generate samples through iterative denoising—we propose a feedback diffusion model (FB-Diffusion). Starting from experimentally validated BBBPs as training data, FB-Diffusion incorporates an external analyzer to provide feedback during generation, enabling diverse sequence generation in small-sample settings. After the convergence of the generator, pseudo-BBBPs are produced and merged with real BBBPs to form the positive class, which is then combined with an equal number of non-BBBPs to construct an augmented training set for training the classifier.
For the classification task, we propose a dual-stream Transformer classifier integrating sequence structure and physicochemical properties. One branch employs self-attention to learn residue-level and motif-level contributions from the peptide sequence, while the other encodes key physicochemical features such as hydrophobic surface area, molecular charge, number of rotatable bonds, and polarizability. 
Experimental results demonstrate that high-quality samples generated by FB-Diffusion significantly alleviate data scarcity and class imbalance. The dual-stream classifier achieves state-of-the-art area under the curve (AUC) and accuracy (ACC) on the test set.

\section{Results and Discussions}\label{sec:results}
\subsection{Comparison of DREAM-$\text{B}^3$P with state-of-the-art methods}
In {\em Experiment 1}, we compare the BBBP identification performance of the proposed DREAM-B$^3$P against several baseline methods: Deep-B$^3$P, B$^3$Pred, BBPpredict and Augur. All models (if publicly available online) were trained on the same training set and evaluated on an identical and independent test set. For methods employing data augmentation, their respective pseudo-BBBPs were incorporated into the positive class of the training set. The test set exclusively consists of real samples (without pseudo-BBBPs) and contains equal numbers of BBBPs and non-BBBPs.

As shown in Table \ref{table:EXP1}, the proposed DREAM-B$^3$P outperforms other algorithms in AUC, ACC, and MCC, achieving improvements of 4.3\%, 17.8\%, and 14.9\%, respectively, over the second-best method. DREAM-B$^3$P also demonstrates balanced sensitivity (SN) and specificity (SP), indicating no substantial bias toward either the positive (BBBPs) or negative (non-BBBPs) class. This balance can be attributed to the high-quality pseudo-BBBPs generated by the proposed FB-Diffusion, which effectively mitigates class imbalance.
In contrast, B$^3$Pred, which does not utilize data augmentation, exhibits a significantly higher SP than SN, suggesting a prediction bias toward non-BBBPs. Deep-B$^3$P, which was augmented with FB-GAN, also tends to classify samples as non-BBBPs, indicating that while FB-GAN produces moderate-quality pseudo-BBBPs, there remains potential for further improvement.
\begin{table*}
	\begin{center}   
		\caption{Performance comparison of DREAM-B$^3$P with existing BBBP prediction methods on the test set.}  
		\label{table:EXP1} 
		\scriptsize
		\renewcommand{\tabularxcolumn}[1]{m{#1}}
		\begin{tabularx}{\textwidth}{>{\centering\arraybackslash}X|>{\centering\arraybackslash}X|>{\centering\arraybackslash}X|>{\centering\arraybackslash}X|>{\centering\arraybackslash}X|>{\centering\arraybackslash}X>{\centering\arraybackslash}X}
			\toprule
			& DREAM-B$^3$P & Deep-B$^3$P & B$^3$Pred & BBPpredict & Augur \\
			
			\midrule
			AUC & \textbf{0.951} & \underline{0.912} & 0.729 & 0.841 & 0.857 \\
			SN  & \textbf{0.912} & \underline{0.774} & 0.227 & 0.773 & 0.745 \\ 
			SP  & 0.858 & \underline{0.905} & \textbf{0.969} & 0.794 & 0.754 \\
			ACC & \textbf{0.886} & \underline{0.826} & 0.598 & 0.784 & 0.754 \\
			MCC & \textbf{0.773} & \underline{0.673} & 0.292 & 0.567 & 0.509 \\
			
			\bottomrule
		\end{tabularx} 
	\end{center} 
	\raisebox{-0.5ex}[0pt][0pt]{\footnotesize{The best results are highlighted with \textbf{bold} and the second best are highlighted with \underline{underline}}.} 
\end{table*}

\subsection{Impact of sequence and physicochemical features}
\begin{figure} [h]
	\centering
	
	\subfigure{
		\begin{minipage}[t]{0.8\linewidth}
			\centering
			\includegraphics[width=1\linewidth]{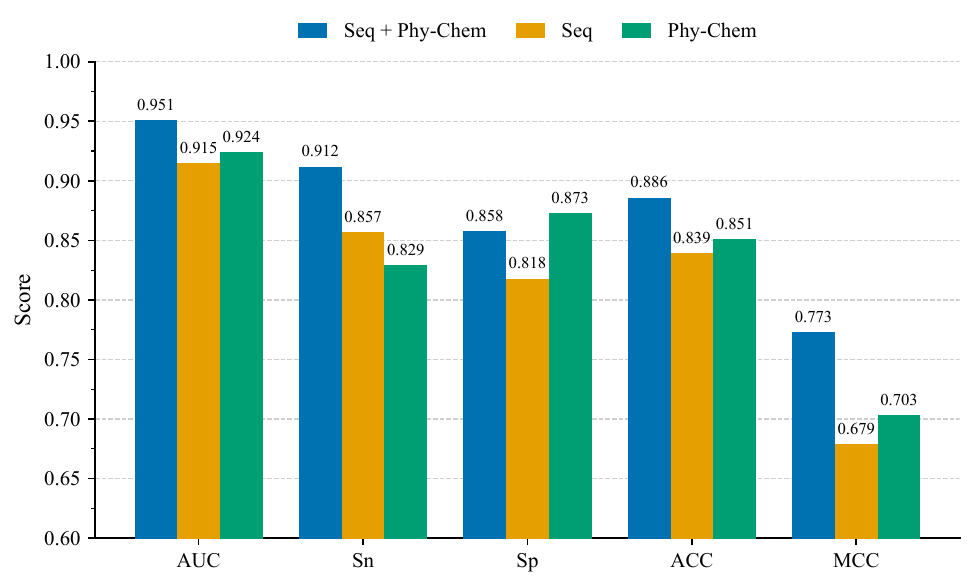}
	\end{minipage}}
	
	\caption{Results of contribution of sequence and physicochemical features.}
	\label{fig:EXP2_plot1}
\end{figure}

In {\em Experiment 2}, we explore the impact of sequence features and physicochemical features on model performance. The training set remains the augmented dataset. We compare three classifier variants: (i) a dual-stream classifier that utilizes both feature types, (ii) a sequence-feature-only classifier (derived from the dual-stream model by disabling the physicochemical feature branch), and (iii) a physicochemical-feature-only classifier (derived by disabling the sequence feature branch).
As illustrated in Fig. \ref{fig:EXP2_plot1}, the sequence-only and physicochemical-only classifiers exhibit slightly different biases toward predicting BBBPs and non-BBBPs. Combining both types of features in the dual-stream model leads to further improvements in AUC, ACC, and MCC, demonstrating the complementary benefits of integrating sequence and physicochemical properties.

\subsection{Impact of data augmentation using feedback diffusion}
\begin{figure} [h]
	\centering
	
	\subfigure{
		\begin{minipage}[t]{0.6\linewidth}
			\centering
			\includegraphics[width=1\linewidth]{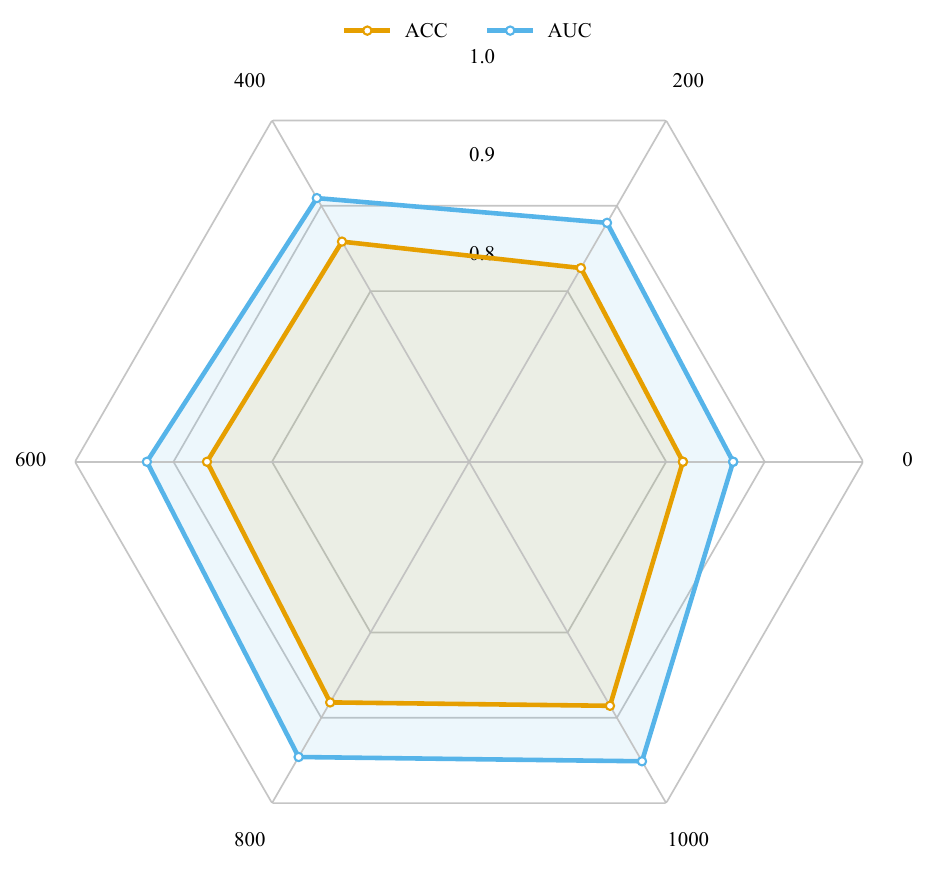}
	\end{minipage}}
	
	\caption{Results for data augmentation performance as the generator trains.}
	\label{fig:EXP3_plot1}
\end{figure}

In {\em Experiment 3}, we evaluate the impact of data augmentation on classifier performance. The FB-Diffusion model was trained for a total of 1000 epochs. At every 200 epochs, we used the current generator to produce 6000 pseudo-BBBPs, added these to the classifier’s training set, and then retrained the dual-stream classifier. Note that the final model at epoch 1000 corresponds to the one used in {\em Experiment 1}. 
Fig. \ref{fig:EXP3_plot1} plots the AUC and ACC of the classifier on the test set as FB-Diffusion training progresses.
We observe that as FB-Diffusion training advances, the classifier’s AUC and ACC gradually improve, plateauing around 800 epochs and peaking at 1000 epochs. This indicates that the generator produces data increasingly resembling real BBBPs, thereby helping the classifier better identify BBBPs.

Furthermore, after training FB-Diffusion for 1000 epochs, we compared the effect of different augmentation intensities—that is, incorporating varying numbers of pseudo-BBBPs into the training set while keeping the total count of BBBPs and pseudo-BBBPs equal to the number of non-BBBPs. As shown in Table \ref{table:EXP3}, models with data augmentation consistently improve performance, with the largest gain observed when using 6000 pseudo-BBBPs. Due to the high quality of pseudo-BBBPs generated by the proposed FB-Diffusion, using more of them enables the classifier to better distinguish BBBPs from non-BBBPs, resulting in higher accuracy.
In addition, we employed the FB-GAN model, proposed in \cite{gupta2019feedback} and adopted in \cite{tang2024deepb3p}, to generate 6000 pseudo-BBBPs for augmentation. Under the same sample size, data augmentation using the proposed FB-Diffusion yields better performance than FB-GAN, further confirming that FB-Diffusion produces higher-quality pseudo-BBBPs.

%
%

\begin{table*}
	\centering
	\caption{Classifier performance with different numbers of augmented pseudo-BBBPs (from FB-Diffusion), and comparison with FB-GAN augmentation.}
	\label{table:EXP3}
	\scriptsize
	\renewcommand{\tabularxcolumn}[1]{m{#1}} 
	
	\begin{tabularx}{\textwidth}{
			>{\centering\arraybackslash}X   
			| >{\centering\arraybackslash}X
			| >{\centering\arraybackslash}X
			| >{\centering\arraybackslash}X
			| >{\centering\arraybackslash}X
			| >{\centering\arraybackslash}X
			|
			>{\centering\arraybackslash}X
		}
		\toprule
		Generator & \multicolumn{5}{c!{\vrule width 0.1pt}}{FB-Diffusion}
		& \multicolumn{1}{c}{FB-GAN} \\
		\midrule
		Number of pseudo-BBBPs & 0 & 1000 & 2000 & 4000 & 6000 & 6000 \\
		\midrule
		AUC & 0.868 & 0.946 & 0.938 & \textbf{0.952} & \underline{0.951} & 0.927 \\
		SN  & 0.660 & 0.888 & 0.883 & \underline{0.888} & \textbf{0.912} & 0.830 \\
		SP  & \textbf{0.992} & 0.824 & 0.839 & 0.829 & 0.858 & \underline{0.882} \\
		ACC & 0.817 & 0.858 & \underline{0.863} & 0.860 & \textbf{0.886} & 0.856 \\
		MCC & 0.682 & 0.716 & \underline{0.725} & 0.721 & \textbf{0.773} & 0.713 \\
		\bottomrule
	\end{tabularx}
	
	\vspace{1ex}
	\footnotesize{The best results are highlighted with \textbf{bold} and the second best are highlighted with \underline{underline}.}
\end{table*}


\section{Conclusion}\label{sec:conclusion}
In this paper, we introduced DREAM-B$^3$P, a deep learning framework that synergizes a feedback diffusion model with a dual-stream Transformer network to tackle two fundamental challenges in BBBP prediction: data imbalance and limited feature representation.
Extensive experimental results confirm that our method achieves state-of-the-art performance across multiple evaluation metrics. In particular, the proposed FB-Diffusion model effectively generates diverse and high-quality pseudo-BBBPs. Incorporating these pseudo-samples significantly enhances the classifier’s ability to discriminate between BBBPs and non-BBBPs on an independent test set.
Moreover, the dual-stream architecture not only yields superior prediction accuracy but also provides interpretability regarding the respective roles of sequence components and physicochemical properties in BBB penetration. 
We also conducted ablation studies and comparative analyses with competing augmentation strategies (e.g., GAN-based approaches), which further validated the stability of samples generated by FB-Diffusion.
Looking forward, the DREAM-B$^3$P framework is not limited to BBBP recognition; it can be extended to other peptide functional classification tasks under small-sample and imbalanced-data scenarios. 

\section{Methods}\label{sec:methods}
\subsection{Benchmark datasets}
In this paper, the BBBPs used to train our models are obtained from \cite{tang2024deepb3p}, which integrates the datasets released with BBPpredict \cite{chen2022bbppredict} together with BBBPs collected from publicly available databases and prediction tools that are not included in BBPpredict datasets, yielding a total of 428 BBBPs and 6950 non-BBBPs. The lengths of the peptide sequences in the datasets do not exceed 50 amino acids (aa).

\begin{figure} [h]
	\centering
	
	\subfigure{
		\begin{minipage}[t]{0.8\linewidth}
			\centering
			\includegraphics[width=1\linewidth]{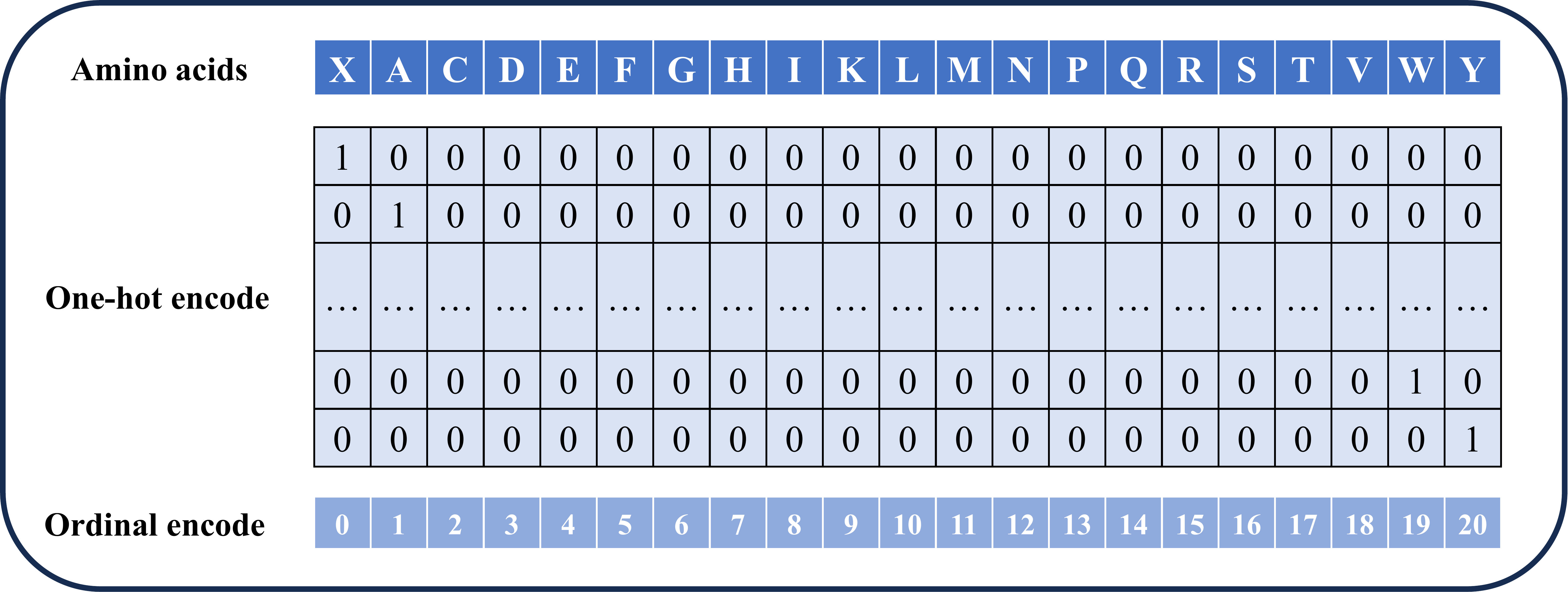}
	\end{minipage}}
	
	\caption{Schematic of the encoding schemes of amino acids.}
	\label{fig:aa}
\end{figure}

For consistency, shorter peptide sequences are padded with a virtual amino acid ``X" at the end so that all sequences are standardized to a length of 50 aa. Only the BBBPs are used to train the generative model (FB-Diffusion). All 428 BBBPs comprise the training data for FB-Diffusion and are represented via one-hot encoding. Specifically, each amino acid (including the 20 standard amino acids and one virtual amino acid) is encoded as a 21-dimensional binary vector (with a single 1 and all other entries 0). The one-hot encoding for each amino acid is illustrated in Fig. \ref{fig:aa}. Therefore, a BBBP of length 50 aa is represented by a $50\times21$ binary matrix.
For the classification model, the training set consists of 378 BBBPs and 6900 non-BBBPs, while the independent test set contains 50 BBBPs and 50 non-BBBPs (all selected at random from the total data). An ordinal encoding scheme is applied to represent the BBBPs. Specifically, each amino acid is represented by an integer from 0 to 20. The ordinal encoding for each amino acid is provided in Fig. \ref{fig:aa}. Thus, each BBBP of length 50 aa is encoded as a $1\times50$ vector. In addition, the classification model leverages physicochemical features of the BBBPs, which are derived from the peptide structures. The details of physicochemical features are provided in Appendix A.

\subsection{Feedback diffusion model for data augmentation}
\begin{figure} [h]
	\centering
	
	\subfigure{
		\begin{minipage}[t]{0.8\linewidth}
			\centering
			\includegraphics[width=1\linewidth]{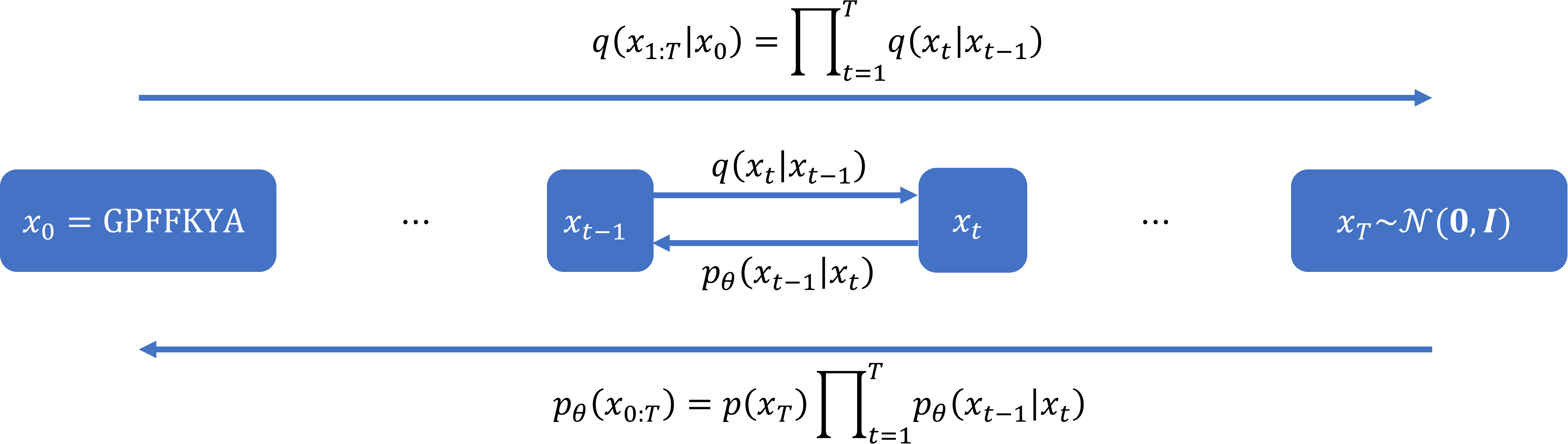}
	\end{minipage}}
	
	\caption{Schematic of the diffusion model.}
	\label{fig:Diffusion}
\end{figure}

Diffusion models are latent variable-based generative frameworks comprising two Markov chain-based processes: a predefined forward process and a learnable reverse process.  
The forward process progressively corrupts clean training samples (BBBPs in this paper) through additive Gaussian noise, transforming the data until the resulting latent variables approximately follow a standard normal distribution. Subsequently, a reverse process is trained via variational inference to reconstruct samples closely resembling the original training data from these latent variables. 
Specifically, denote real BBBPs and the latent variable in the $t$-th forward Markov chain by $\m{x}_0$ and $\m{x}_t, t=1,...,T$, respectively. 
Let the last latent variable $\m{x}_T$ follow a standard normal distribution, i.e., $p\sbra{\m{x}_T}=\cN\sbra{\m{x}_T; \m{0}, \m{I}}$.
The $t$-th forward Markov chain $q\sbra{\m{x}_t|\m{x}_{t-1}}$ continuously corrupts latent variable $\m{x}_{t-1}$ until $q\sbra{\m{x}_T|\m{x}_{T-1}}$ approximates $p\sbra{\m{x}_T}$, resulting in the forward process $q\sbra{\m{x}_{1:T}|\m{x}_0}
=\prod_{t=1}^{T}q\sbra{\m{x}_t|\m{x}_{t-1}}$. 
In contrast, the $t$-th reverse Markov chain $p_{\theta}\sbra{\m{x}_{t-1}|\m{x}_t}$ progressively denoises $\m{x}_t$ until the reconstructed sample approximates ground truth $\m{x}_0$, forming the reverse process $p_{\theta}\sbra{\m{x}_{0:T}}=p\sbra{\m{x}_T} \prod_{t=1}^{T}p_{\theta}\sbra{\m{x}_{t-1}|\m{x}_t}$, where $\theta$ is a learnable parameter.
To achieve the above objective, the diffusion model aims to maximize the log-likelihood $p_{\theta}\sbra{\m{x}_0}$, which can be relaxed to maximize the evidence lower bound (ELBO), denoted by $\bE_{q\sbra{\m{x}_{1:T}|\m{x}_0}}\mbra{\log\dfrac{p_{\theta}\sbra{\m{x}_{0:T}}}{q\sbra{\m{x}_{1:T}|\m{x}_0}}}$.
The diffusion model is illustrated in Fig. \ref{fig:Diffusion}, and the detailed algorithm derivation can be found in \cite{luo2022understanding}.

\begin{figure} [h]
	\centering
	
	\subfigure{
		\begin{minipage}[t]{0.8\linewidth}
			\centering
			\includegraphics[width=1\linewidth]{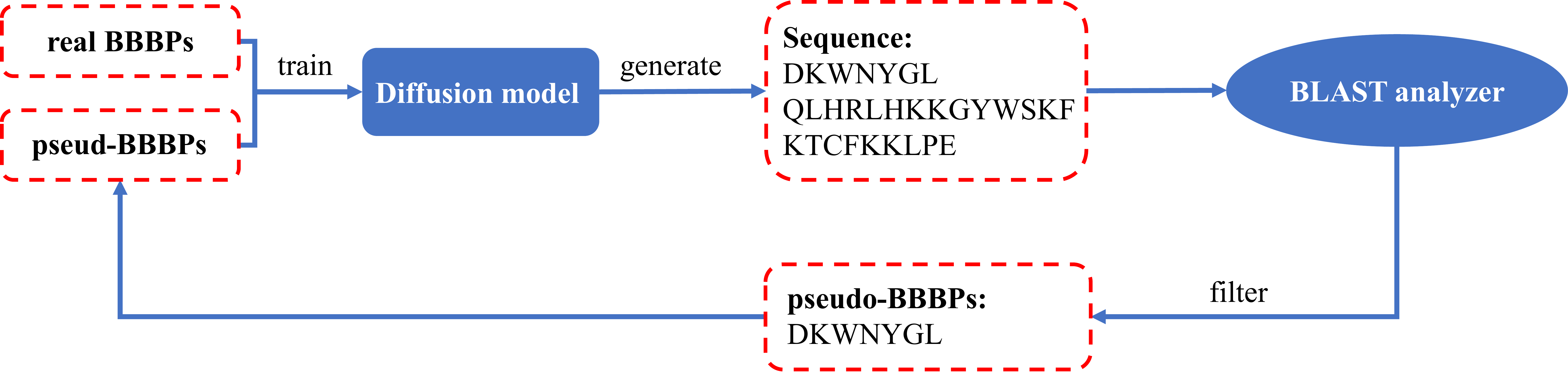}
	\end{minipage}}
	
	\caption{Schematic of FB-Diffusion.}
	\label{fig:FBDiffusion}
\end{figure}

Due to the scarcity of experimentally verified BBBPs, the diffusion model’s training is prone to underfitting; that is, the distribution of samples reconstructed by the reverse diffusion process can remain a non-trivial distance away from the true BBBP distribution. To mitigate this issue, we adopt the feedback mechanism proposed in \cite{gupta2019feedback}, which connects an independent BLAST analyzer to the diffusion model to evaluate the similarity between generated samples and the real BBBPs.
Specifically, the diffusion model is first pretrained using the real BBBPs so that it can produce valid candidate sequences. The feedback mechanism is then activated: in each subsequent training epoch, the generator produces a batch of samples that are input into the analyzer. The analyzer assigns a score to each sample, and those exceeding a predefined threshold are retained as pseudo-BBBPs. These high-scoring pseudo-BBBPs are added to the real BBBPs and the diffusion model is retrained on this expanded set. As training continues, the oldest pseudo-BBBPs are periodically replaced by newly generated high-scoring samples, thereby progressively refining the generator’s training data with higher-quality samples. The FB-Diffusion process is illustrated in Fig. \ref{fig:FBDiffusion}.

\subsection{Dual-stream Transformer network for identifying BBBPs}
Our classifier comprises two independent Transformer encoders to extract complementary features: one encoder processes the peptide sequence (to capture sequence motifs and structural patterns), and the other processes a vector of physicochemical properties.
The outputs of these two encoders are concatenated and passed through a multilayer perceptron (MLP) to predict whether the peptide can penetrate the BBB.
The core of the Transformer encoder is the multi-head self-attention mechanism, which for each amino acid in the sequence computes attention weights over all amino acids in the peptide (including itself) and aggregates these into a contextualized representation.
Formally, let the input of the $l$-th encoder block be $\m{X}_{l-1}\in\bR^{d_{\text{input}}\times d_{\text{model}}}$. We first add a sinusoidal positional encoding $\m{P}\in\bR^{d_{\text{input}}\times d_{\text{model}}}$ to this input to obtain $\widetilde{\m{X}}_{l-1}=\m{X}_{l-1}+\m{P}$.
For each layer $l = 1,2,\dots,L$ and each attention head $h = 1,2,\dots,H$, we compute the queries, keys, and values via learned linear projections:
\begin{equation}
	\begin{split}
		\m{Q}_{l,h} 
		= \widetilde{\m{X}}_{l-1} \m{W}_{l,h}^Q,~~
		\m{K}_{l,h} 
		= \widetilde{\m{X}}_{l-1} \m{W}_{l,h}^K,~~
		\m{V}_{l,h} 
		= \widetilde{\m{X}}_{l-1} \m{W}_{l,h}^V,
	\end{split}
\end{equation}
where $\m{W}_{l,h}^Q, \m{W}_{l,h}^K\in\bR^{d_{\text{model}}\times d_k}$ and $\m{W}_{l,h}^V\in\bR^{d_{\text{model}}\times d_v}$ are projection matrices. 
We then perform scaled dot-product attention:
\begin{equation}
	\text{Attn}_{l,h}\sbra{\m{X}_{l-1}} 
	= \text{softmax}\sbra{\dfrac{\m{Q}_{l,h}\m{K}_{l,h}^T}{\sqrt{d_k}}}\m{V}_{l,h}.
\end{equation}
The outputs of the $H$ heads are concatenated and projected back to the model dimension:
\begin{equation}
	\text{MultiHead}_l\sbra{\m{X}_{l-1}}
	= \text{Concat}\sbra{\text{Attn}_{l,1},\cdots,\text{Attn}_{l,H}}\m{W}_l^O,
\end{equation}
with $\m{W}_l^O\in\bR^{Hd_v\times d_{\text{model}}}$.

Each encoder block applies the multi-head attention followed by a position-wise feed-forward network (FFN), with residual connections and layer normalization (LN) at each layer, producing:
\begin{align}
	\m{Y}_l
	& = \text{LN}\sbra{\m{X}_{l-1}+\text{MultiHead}_l\sbra{\m{X}_{l-1}}}, \\
	\m{X}_l
	& = \text{LN}\sbra{\m{Y}_l+\text{FFN}\sbra{\m{Y}_l}}.
\end{align}
The final output of the Transformer encoder is $\m{X}_L$ (the output of the last layer).

\begin{figure} [h]
	\centering
	
	\subfigure{
		\begin{minipage}[t]{0.95\linewidth}
			\centering
			\includegraphics[width=1\linewidth]{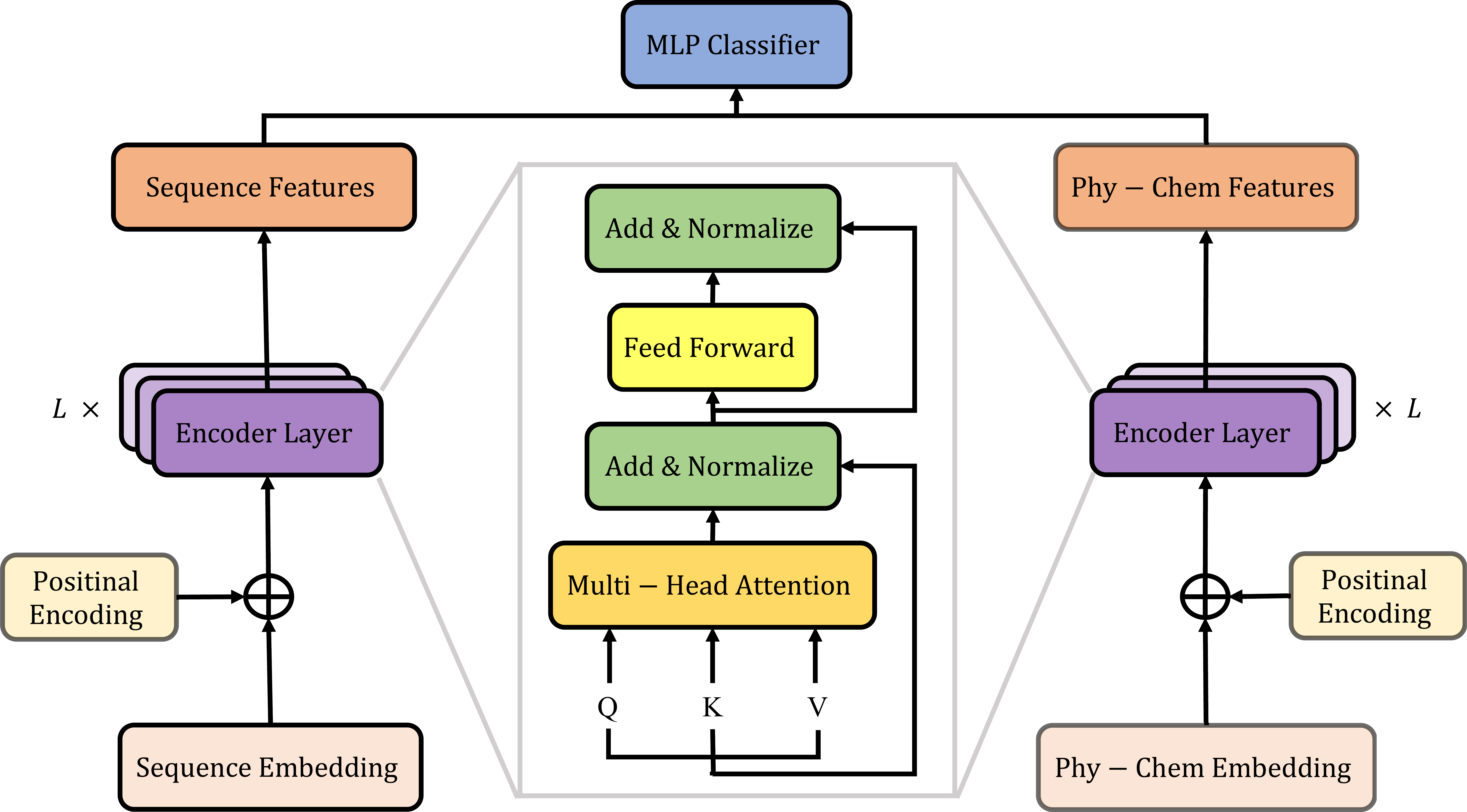}
	\end{minipage}}
	
	\caption{Schematic of the dual Transformer classifier.}
	\label{fig:dual-cls}
\end{figure}

We employ two such Transformer encoders in parallel: one yields a sequence feature representation $\m{X}_L^{\text{Seq}}$ and the other yields a physicochemical feature representation $\m{X}_L^{\text{Phy-Chem}}$.
We then concatenate these $\mbra{\m{X}_L^{\text{Seq}}, \m{X}_L^{\text{Phy-Chem}}}$ and feed them into an MLP, which maps the integrated features to the target classes through fully connected layers with nonlinear activation functions, thereby achieving sequence classification. The detailed dual classifier is illustrated in Fig. \ref{fig:dual-cls}.

\begin{figure} [h]
	\centering
	
	\subfigure{
		\begin{minipage}[t]{0.8\linewidth}
			\centering
			\includegraphics[width=1\linewidth]{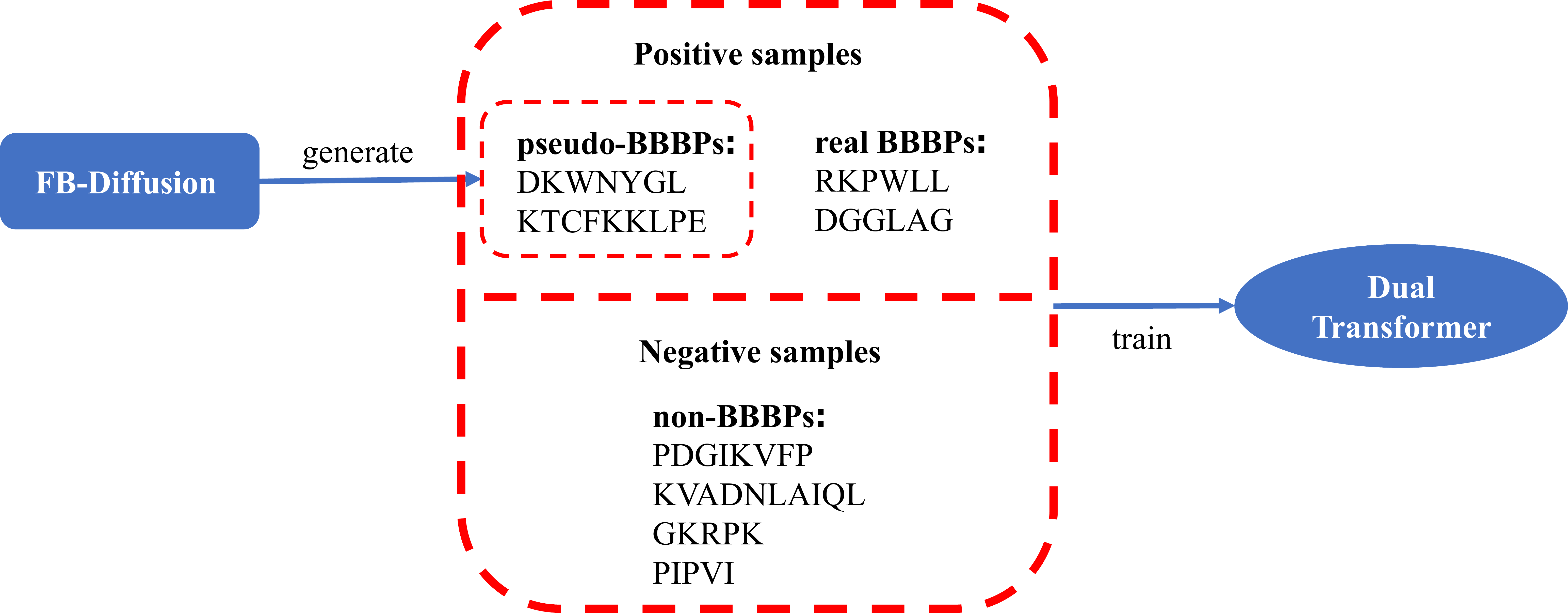}
	\end{minipage}}
	
	\caption{Schematic of DREAM-B$^3$P.}
	\label{fig:DREAM-B3P}
\end{figure}

To address the class imbalance between BBBPs and non-BBBPs in training, we augment the positive class with the pseudo-BBBPs generated by FB-Diffusion.
In other words, we add the FB-Diffusion sequences to the BBBP training set before training the classifier. 
This integrated approach is termed the dual-stream Transformer network enhanced by a feedback diffusion model for blood-brain barrier penetrating peptide prediction (DREAM-B$^3$P), as depicted in Fig. \ref{fig:DREAM-B3P}.

\subsection{Training process}
The overall training procedure consists of two phases: generator training followed by classifier training.
For the generator, we used the Adam optimizer with a learning rate of $1\times10^{-4}$ and a batch size of 64, and trained for 1000 epochs. At every 200 epochs, the current generator produced a batch of 6000 pseudo-BBBPs, which were subsequently screened by the BLAST analyzer. The validated sequences were then added to the positive subset of the training set and used to retrain the generator in the subsequent epochs.
For the classifier, two independent Transformer encoders were employed with $d_{\text{model}}=512$ and $d_k=d_v=32$. To mitigate overfitting, we implemented a 10-fold cross-validation strategy combined with an early stopping mechanism. The training used a batch size of 64, with each fold trained for up to 200 epochs. This process yielded ten independently trained models and the final prediction was obtained by averaging the predictions from all these models.

\subsection{Performance evaluation metrics}
We evaluated model performance using several standard metrics: accuracy (ACC), sensitivity (SN), specificity (SP), and Matthews correlation coefficient (MCC). These are defined as follows:
\begin{equation}
	\begin{split}
		\text{ACC} 
		& = \dfrac{n_\text{TP}+n_\text{TN}}{n_\text{TP}+n_\text{TN}+n_\text{FP}+n_\text{FN}}, \\
		\text{SN}
		& = \dfrac{n_\text{TP}}{n_\text{TP}+n_\text{FN}}, \\
		\text{SP}
		& = \dfrac{n_\text{TN}}{n_\text{TN}+n_\text{FP}}, \\
		\text{MCC} 
		& = \dfrac{n_\text{TP}n_\text{TN}-n_\text{FP}n_\text{FN}}{\sqrt{\sbra{n_\text{TP}+n_\text{FP}}\sbra{n_\text{TP}+n_\text{FN}}\sbra{n_\text{TN}+n_\text{FP}}\sbra{n_\text{TN}+n_\text{FN}}}},
	\end{split}
\end{equation}
where $n_{\mathrm{TP}}$, $n_{\mathrm{TN}}$, $n_{\mathrm{FP}}$, and $n_{\mathrm{FN}}$ denote the numbers of true positives, true negatives, false positives, and false negatives, respectively (with “positive” referring to BBBPs and “negative” to non-BBBPs). 
We also report the area under the receiver operating characteristic curve (AUC). AUC ranges from 0.5 (no better than random guessing) to 1 (perfect classification), with higher values indicating better discriminatory ability.

%
%
%

%

\begin{appendices}
%
%

\section{Feature Principles and Formulas}
This appendix details the principles and mathematical formulas of the 14 physicochemical features used for biological sequence feature extraction.

\subsection{Amino Acid Composition (AAC)}
AAC describes the frequency of each amino acid in a sequence. For a sequence \( S \) with length \( L \) (after removing gaps), the frequency of amino acid \( a \) (where \( a \in \{A, C, D, \dots, Y\} \), 20 standard amino acids) is calculated as:
\[
\text{AAC}(a) = \frac{\text{Count}(a, S)}{L}
\]
where \( \text{Count}(a, S) \) is the number of occurrences of amino acid \( a \) in \( S \).

\subsection{Amphiphilic Pseudo Amino Acid Composition (APAAC)}
APAAC combines amino acid composition with sequence-order information based on amphiphilic properties. It is defined as:
\[
\text{APAAC}_i = 
\begin{cases} 
	\frac{\text{Count}(a_i, S)}{L + \lambda \sum_{k=1}^\lambda \theta_k} & (i = 1, 2, \dots, 20) \\
	\frac{\lambda \theta_{i-20}}{L + \lambda \sum_{k=1}^\lambda \theta_k} & (i = 21, 22, \dots, 20 + \lambda)
\end{cases}
\]
where \( \lambda \) is the order of sequence correlation (set to 2 in the code), \( \theta_k \) measures the correlation between residues separated by \( k \) positions, calculated using normalized physicochemical property values, \( w \) is a weight factor (set to 0.05), \( L \) is the sequence length (after gap removal).

\subsection{Amino Acid Substitution Matrix Distance (ASDC)}
ASDC quantifies the frequency of all possible amino acid pairs (including non-consecutive) in the sequence. For a sequence \( S \) of length \( L \), the frequency of amino acid pair \( (a, b) \) is:
\[
\text{ASDC}(a, b) = \frac{\sum_{1 \leq i < j \leq L} \mathbb{I}(S[i] = a \land S[j] = b)}{\sum_{1 \leq i < j \leq L} 1}
\]
where \( \mathbb{I}(\cdot) \) is the indicator function (1 if true, 0 otherwise), and the denominator is the total number of residue pairs (\( \frac{L(L-1)}{2} \)).

\subsection{Composition of k-Spaced Amino Acid Pairs (CKSAAP)}
CKSAAP captures the frequency of amino acid pairs separated by \( k \) residues. For a given gap \( k \) (ranging from 0 to 3 in the code), the frequency of pair \( (a, b) \) is:
\[
\text{CKSAAP}(a, b, k) = \frac{\sum_{i=1}^{L - k - 1} \mathbb{I}(S[i] = a \land S[i + k + 1] = b)}{L - k - 1}
\]
where \( S[i] \) denotes the \( i \)-th residue in \( S \), and \( L \) is the sequence length (after gap removal).

\subsection{Composition-Transition-Distribution (CTD)}
CTD is a set of features based on physicochemical properties of amino acids, divided into three sub-features:

\subsubsection{Composition (CTDC)}
CTDC measures the proportion of residues belonging to each of three physicochemical classes for a given property. For class \( c \in \{G1, G2, G3\} \):
\[
\text{CTDC}(c) = \frac{\text{Number of residues in class } c}{L}
\]

\subsubsection{Transition (CTDT)}
CTDT measures the frequency of transitions between different classes. For classes \( c_1 \) and \( c_2 \):
\[
\text{CTDT}(c_1, c_2) = \frac{\text{Number of transitions from } c_1 \text{ to } c_2 \text{ (or vice versa)}}{L - 1}
\]

\subsubsection{Distribution (CTDD)}
CTDD describes the position distribution of residues in each class, characterized by the normalized positions of the 1st, 25\%, 50\%, 75\%, and last occurrence of residues in class \( c \):
\[
\text{CTDD}(c, p) = \frac{\text{Position of the } p\text{-th percentile residue in class } c}{L} \times 100
\]
where \( p \in \{1, 25, 50, 75, 100\} \).

\subsection{Dipeptide Distance Distribution (DDE)}
DDE characterizes the normalized frequency of dipeptides, adjusted by their expected frequency based on codon usage. For dipeptide \( ab \):
\[
\text{DDE}(ab) = \frac{f(ab) - T_m(ab)}{\sqrt{T_v(ab)}}
\]
where:
- \( f(ab) = \frac{\text{Count}(ab, S)}{L - 1} \) is the observed frequency of \( ab \),
- \( T_m(ab) = \left( \frac{\text{Codons}(a)}{61} \right) \times \left( \frac{\text{Codons}(b)}{61} \right) \) is the expected frequency based on codon counts,
- \( T_v(ab) = T_m(ab) \times (1 - T_m(ab)) / (L - 1) \) is the variance term.

\subsection{Dipeptide Composition (DPC)}
DPC calculates the frequency of each consecutive dipeptide (two-residue pair) in the sequence. For dipeptide \( ab \) (where \( a, b \in \{A, C, \dots, Y\} \)):
\[
\text{DPC}(ab) = \frac{\text{Count}(ab, S)}{L - 1}
\]
where \( \text{Count}(ab, S) \) is the number of times \( ab \) appears consecutively in \( S \), and \( L \) is the sequence length (after gap removal).

\subsection{Tripeptide Composition (TPC)}
TPC is the frequency of each consecutive tripeptide (three-residue pair) in the sequence. For tripeptide \( abc \):
\[
\text{TPC}(abc) = \frac{\text{Count}(abc, S)}{L - 2}
\]
where \( \text{Count}(abc, S) \) is the number of times \( abc \) appears consecutively in \( S \), and \( L \) is the sequence length (after gap removal).

\subsection{Sequence Pattern (SEP)}
SEP measures the diversity of short subsequences (patterns) of varying lengths. For pattern length \( i \), the feature is:
\[
\text{SEP}(i) = \frac{\text{Number of unique patterns of length } i}{L}
\]
where unique patterns are distinct subsequences \( S[j:j+i] \) for \( j = 1, 2, \dots, L - i + 1 \).

\subsection{Sequence Repeat (SER)}
SER counts the number of consecutive identical residue pairs in the sequence:
\[
\text{SER} = \sum_{i=1}^{L - 1} \mathbb{I}(S[i] = S[i + 1])
\]
where \( \mathbb{I}(\cdot) \) is the indicator function (1 if true, 0 otherwise).

\subsection{Quasi-Sequence Order (QSO)}
QSO integrates amino acid composition with sequence-order information using residue distance matrices (Schneider-Wrede and Grantham). For each matrix type, the features are:
\[
\text{QSO}_{\text{comp}}(a) = \frac{\text{Count}(a, S)}{1 + w \sum_{n=1}^{\lambda} \theta_n}
\]
\[
\text{QSO}_{\text{order}}(n) = \frac{w \cdot \theta_n}{1 + w \sum_{n=1}^{\lambda} \theta_n}
\]
where:
- \( \theta_n = \sum_{i=1}^{L - n} d(S[i], S[i + n])^2 \) ( \( d(\cdot, \cdot) \) is the distance from the matrix),
- \( \lambda \) is the maximum lag (set to 2),
- \( w \) is a weight factor (set to 0.05).

\subsection{Sequence Entropy (SE)}
SE quantifies the uncertainty in amino acid distribution, based on Shannon entropy:
\[
\text{SE} = -\sum_{a \in \text{AA}} p(a) \log_2 p(a)
\]
where \( p(a) = \frac{\text{Count}(a, S)}{L} \) is the frequency of amino acid \( a \), and the sum excludes \( a \) with \( p(a) = 0 \).

\subsection{Sum of Consecutive Nucleotide (Amino Acid) Values (SOCN)}
SOCN calculates the sum of products of consecutive residues, where each residue is mapped to a numerical value based on its ASCII code:
\[
\text{SOCN} = \sum_{i=1}^{L - 1} (v(S[i]) \times v(S[i + 1]))
\]
where \( v(a) = \text{ord}(a) - \text{ord}('A') + 1 \), and \( \text{ord}(\cdot) \) is the ASCII value function.

\subsection{List of 33 Physicochemical Features (Calculated via RDKit)}

The following table presents 33 physicochemical features, all computed using the \texttt{Descriptors} module of the RDKit cheminformatics library. These features quantify key molecular properties of peptide sequences (converted to SMILES format).

\begin{table}[h]
	\centering
	\begin{tabular}{|c|>{\raggedright\arraybackslash}m{2.5cm}|c|>{\raggedright\arraybackslash}m{2.5cm}|c|>{\raggedright\arraybackslash}m{2.5cm}|}
		\hline
		\textbf{No.} & \textbf{Feature Name} & \textbf{No.} & \textbf{Feature Name} & \textbf{No.} & \textbf{Feature Name} \\
		\hline
		1  & Molecular Weight (MolWt) & 12 & Total Ring Count & 23 & Average EState Index \\
		\hline
		2  & Heavy Atom Molecular Weight & 13 & Heavy Atom Count & 24 & Balaban J Index \\
		\hline
		3  & Molar Refractivity (MolMR) & 14 & Number of Carbon Atoms & 25 & Bertz CT Complexity \\
		\hline
		4  & Molecular LogP (MolLogP) & 15 & Number of Oxygen Atoms & 26 & Topological Charge Index \\
		\hline
		5  & Topological Polar Surface Area (TPSA) & 16 & Number of Nitrogen Atoms & 27 & Kier Shape Index \\
		\hline
		6  & Number of Hydrogen Bond Donors & 17 & Number of Sulfur Atoms & 28 & Molecular Volume \\
		\hline
		7  & Number of Hydrogen Bond Acceptors & 18 & Maximum Partial Charge & 29 & Polarizability \\
		\hline
		8  & Number of Rotatable Bonds & 19 & Minimum Partial Charge & 30 & Isoelectric Point (pI) \\
		\hline
		9  & Number of Aromatic Rings & 20 & Average Partial Charge & 31 & Molecular Charge \\
		\hline
		10 & Number of Saturated Rings & 21 & Maximum EState Index & 32 & Fraction of Aromatic Carbons \\
		\hline
		11 & Number of Heterocyclic Rings & 22 & Minimum EState Index & 33 & Hydrophobic Surface Area \\
		\hline
	\end{tabular}
	\caption{33 Physicochemical Features Calculated Using RDKit }
	\label{tab:33_features}
\end{table}

\end{appendices}

\bibliography{myReferences}

\end{document}